\documentclass[aps,chaos,preprint,superscriptaddress,showpacs]{revtex4}

\usepackage{graphicx}% Include figure files
\usepackage{dcolumn}% Align table columns on decimal point
\usepackage{bm}% bold math
\usepackage{amssymb}

\begin{document}

%\preprint{}

\title{Synchronization of chaotic oscillator time scales}


\author{Alexander~E.~Hramov}
\email{aeh@cas.ssu.runnet.ru}
\author{Alexey~A.~Koronovskii}
\author{Yurij~I.~Levin}
%\email{alkor@cas.ssu.runnet.ru}
%\homepage[http:\\]{cas.ssu.runnet.ru}
\affiliation{Department of Nonlinear Processes, Saratov State
University, Astrakhanskaya, 83, Saratov, 410012, Russia}

%\thanks{}


\date{\today}

\begin{abstract}
This paper deals with the chaotic oscillator synchronization. A
new approach to detect the synchronized behaviour of chaotic
oscillators has been proposed. This approach is based on the
analysis of different time scales in the time series generated by
the coupled chaotic oscillators. It has been shown that complete
synchronization, phase synchronization, lag synchronization and
generalized synchronization are the particular cases of the
synchronized behavior called as ``time--scale synchronization''.
The quantitative measure of chaotic oscillator synchronous
behavior has been proposed. This approach has been applied for the
coupled R\"ossler systems.
\end{abstract}


\maketitle


\section*{Introduction}
\label{intro} Synchronization of chaotic oscillators is one of the
fundamental phenomena of nonlinear dynamics. It takes place in
many
physical~\cite{Parlitz:1996_PhaseSynchroExperimental,%
Tang:1998_PhaseSynchroLasers,%
Allaria:2001_PhaseSynchroLaser,Ticos:2000_PlasmaDischarge,%
Rosa:2000_PlasmaDischarge,aeh:2003_SynchroDistrSyst} and
biological~\cite{Tass:1998_NeuroSynchro,Anishchenko:2000_humanSynchro,%
Prokhorov:2003_HumanSynchroPRE} processes. It seems to play an
important role in the ability of biological oscillators, such as
neurons, to act
cooperatively~\cite{Elson:1998_NeronSynchro,Rulkov:2002_2DMap,%
Tass:2003_NeuroSynchro}.

There are several different types of synchronization of coupled
chaotic oscillators which have been described theoretically and
observed experimentally~\cite{Pikovsky:2002_SynhroBook,%
Anshchenko:2001_SynhroBook,Pikovsky:2000_SynchroReview,%
Anishchenko:2002_SynchroEng}. The \textit{complete
synchronization} (CS) implies coincidence of states of coupled
oscillators $\mathbf{x}_1(t)\cong\mathbf{x}_2(t)$, the difference
between state vectors of coupled systems converges to zero in the
limit $t\rightarrow\infty$,
\cite{Pecora:1990_ChaosSynchro,%
Pecora:1991_ChaosSynchro,Murali:1994_SynchroIdenticalSyst,%
Murali:1993_SignalTransmission}. It appears when interacting
systems are identical. If the parameters of coupled chaotic
oscillators slightly mismathch, the state vectors are close
$|\mathbf{x}_1(t)-\mathbf{x}_2(t)|\approx 0$, but differ from each
other. Another type of synchronized behavior of coupled chaotic
oscillators wihth slightly mismatched parameters is the
\textit{lag synchronization} (LS), when shifted in time, the state
vectors coincide with each other,
$\mathbf{x}_1(t+\tau)=\mathbf{x}_2(t)$. When the coupling between
oscillator increases the time lag $\tau$ decreases and the
synchronization regime tends to be CS
one~\cite{Rosenblum:1997_LagSynchro,Zhigang:2000_GSversusPS,%
Taherion:1999_LagSynchro}. The \textit{generalized
synchronization} (GS)
\cite{Rulkov:1995_GeneralSynchro,Kocarev:1996_GS,%
Pyragas:1996_WeakAndStrongSynchro} introduced for drive--responce
systems, means that there is some functional relation between
coupled chaotic oscillators, i.e.
$\mathbf{x}_2(t)=\mathbf{F}[\mathbf{x}_1(t)]$.

Finally, it is necessary to mention the \textit{phase
synchronization} (PS) regime. To describe the phase
synchronization the instantaneous phase $\phi(t)$ of a chaotic
continuous time series is usually
introduced \cite{Pikovsky:2000_SynchroReview,Anishchenko:2002_SynchroEng,%
Pikovsky:2002_SynhroBook,Anshchenko:2001_SynhroBook,%
Rosenblum:1996_PhaseSynchro,Osipov:1997_PhaseSynchro}. The phase
synchronization means the entrainment of phases of chaotic
signals, whereas their amplitudes remain chaotic and uncorrelated.

All synchronization types mentioned above are associated with each
other (see for detail~\cite{Parlitz:1996_PhaseSynchroExperimental,%
Rulkov:1995_GeneralSynchro,Zhigang:2000_GSversusPS}), but the
relationship between them is not completely clarified yet. For
each type of synchronization there are their own ways to detect
the synchronized behavior of coupled chaotic oscillators. The
complete synchronization can be displayed by means of comparison
of system state vectors $\mathbf{x}_1(t)$ and $\mathbf{x}_2(t)$,
whereas the lag synchronization can be determined by means of the
similarity function~\cite{Rosenblum:1997_LagSynchro}. The case of
the generalized synchronization is more intricate because the
functional relation $\mathbf{F}[\cdot]$ can be very complicated,
but there are several methods to detect the synchronized behavior
of coupled chaotic oscillators, such as the auxiliary system
approach~\cite{Rulkov:1996_AuxiliarySystem} or the method of
nearest
neighbors~\cite{Rulkov:1995_GeneralSynchro,Pecora:1995_statistics}.

Finally, the phase synchronization of two coupled chaotic
oscillators occurs if the difference between the instantaneous
phases $\phi(t)$ of chaotic signals $\mathbf{x}_{1,2}(t)$ is
bounded by some constant
\begin{equation}
|\phi_{1}(t)-\phi_{2}(t)|<\mathrm{const}. \label{eq:PhaseLocking}
\end{equation}
It is possible to define a mean frequency of chaotic signal
\begin{equation}
\bar{\Omega}=\lim\limits_{t\rightarrow\infty}\frac{\phi(t)}{t}=
\langle\dot{\phi}(t)\rangle, \label{eq:MeanFrequency}
\end{equation}
which should be the same for both coupled chaotic systems, i.e.,
the phase locking leads to the frequency entrainment. It is
important to notice, to obtain correct results the mean frequency
$\bar{\Omega}$ of chaotic signal $\mathbf{x}(t)$ should coincide
with the main frequency $\Omega_0=2\pi f_0$ of the Fourier
spectrum (for detail, see~\cite{Anishchenko:2004_ChaosSynchro}).
There is no general way to introduce the phase for chaotic time
series. There are several approaches which allow to define the
phase for ``good'' systems with simple topology of chaotic
attractor (so--called ``phase coherent attractor''), the Fourier
spectrum of which contains the single main frequency $f_0$.

First of all, the plane in the system phase space may exist, on
which the projection of the chaotic attractor looks like circular
band. For such plane the coordinates $x$ and $y$ can be
introduced, the origin of this coordinate system should be placed
somewhere near the center of the chaotic attractor projection. In
this case the phase can be introduced as an angle in considered
coordinate
system~\cite{Pikovsky:1997PhaseSynchro,Rosenblum:1997_LagSynchro},
but for that all trajectories of chaotic attractor projection on
the $(x,y)$--plane should revolve around the origin. Sometimes, a
coordinate transformation can be used to obtain a proper
projection~\cite{Pikovsky:1997PhaseSynchro,Pikovsky:2002_SynhroBook}.
One can also use the velocities $\dot x$ and $\dot y$, if the
projections of chaotic trajectories on the plane $(\dot x,\dot y)$
always rotate around the origin and in some cases this approach is
more
suitable~\cite{Rosenblum:2002_FrequencyMeasurement,Osipov:2003_3TypesTransitions}.
Another way to define the phase $\phi(t)$ of chaotic time series
$x(t)$ is the constructing of the analytical
signal~\cite{Pikovsky:2000_SynchroReview,Rosenblum:1996_PhaseSynchro}
using the Hilbert transform. Moreover, the Poincar\'e secant
surface can be used for the introducing of the instantaneous phase
of chaotic dynamical
system~\cite{Pikovsky:2000_SynchroReview,Rosenblum:1996_PhaseSynchro}.
Finally, the phase of chaotic time series can be introduced by
means of the continuous wavelet
transform~\cite{Lachaux:2000_WVTSynchro}, but the appropriate
wavelet function and its parameters should be
chosen~\cite{Quiroga:2002_Kraskov}.

All these approaches give correct results for ``good'' systems
with well--defined phase, but fail for oscillators with
non-revolving trajectories. Such chaotic oscillators are often
called as ``systems with ill--defined phase''. The phase
introducing based on the approaches mentioned above for the system
with ill--defined phase leads usually to incorrect
results~\cite{Anishchenko:2004_ChaosSynchro}. Therefore, the phase
synchronization of such systems can be usually detected by means
of the indirect
indications~\cite{Pikovsky:1997PhaseSynchro,Pikovsky:1996_EurophysLett}
and measurements~\cite{Rosenblum:2002_FrequencyMeasurement}.

In this paper we propose a new approach to detect the
synchronization between two coupled chaotic oscillators. The main
idea of this approach consists in the analysis of the system
behavior on different time scales that allows us to consider
different cases of synchronization from the universal positions
\cite{aeh:2004:JETPengl}.
Using the continuous wavelet transform%
~\cite{alkor:2003_WVTBookEng,Daubechies:1992_WVTBook,Kaiser:1994_Wvt,%
Torresani:1995_WVT} we introduce the continuous set of time scales
$s$ and associated with them instantaneous phases $\phi_s(t)$. (In
other words, $\phi_s(t)$ is continuous function of time $t$ and
time scale $s$). As we will show further, if two chaotic
oscillators demonstrate any type of synchronized behavior
mentioned above (CS, LS, PS or GS), in the time series
$\mathbf{x}_{1,2}(t)$ generated by these systems there are time
scales $s$ necessarily correlated for which the phase locking
condition
\begin{equation}
|\phi_{s1}(t)-\phi_{s2}(t)|<\mathrm{const}
\label{eq:SPhaseLocking}
\end{equation}
is satisfied. In other words, CS, LS, PS and GS are the particular
cases of the synchronous coupled chaotic oscillator behavior
called as \textit{``time--scale synchronization''} (TSS).

The structure of this paper is the following. In
section~\ref{Sct:WVTTrans} we discuss the continuous wavelet
transform and the method of the time scales $s$ and associated
with them phases $\phi_s(t)$ definition. In
section~\ref{Sct:PSSynchro} we consider the case of the phase
synchronization of two mutually coupled R\"ossler systems. We
demonstrate the application of our method and discuss the
relationship between our and traditional approaches.
Section~\ref{Sct:IllPhase} deals with the synchronization of two
mutually coupled R\"ossler systems with funnel attractors. In this
case the traditional methods for phase introducing fail and there
is no possibility to detect the phase synchronization regime,
respectively. The synchronization between systems can be revealed
here only by means of the indirect measurements (see for
detail~\cite{Rosenblum:2002_FrequencyMeasurement}). We demonstrate
the efficiency of our method for such cases and discuss the
correlation between PS, LS and CS. In section~\ref{Sct:GSSynchro}
we consider application of our method for the unidirectional
coupled R\"ossler systems when the generalized synchronization is
observed. The quantitative measure of synchronization is described
in section~\ref{Sct:Measure}. The final conclusion is presented in
section~\ref{Sct:Conclusion}.

\section{Continuous wavelet transform}
\label{Sct:WVTTrans}

The continuous wavelet transform is the powerful tool for the
analysis of nonlinear dynamical system behavior. In particular,
the continuous wavelet analysis has been used for the detection of
synchronization of chaotic oscillations in the
brain~\cite{Lachaux:2000_WVTSynchro,Lachaux:2002_BrainCoherence,%
Quyen:2001_WVTvsHilbert} and chaotic laser
array~\cite{DeShazer:2001_WVT_LaserArray}. It has also been used
to detect the main frequency of the oscillations in nephron
autoregulation~\cite{Sosnovtseva:2002_Wvt}. We propose to analyze
the dynamics of coupled chaotic oscillators using the
consideration of system behavior on different time scales $s$ each
of them is characterized by means of its own phase $\phi_s(t)$,
respectively. So, in order to define {\it the continuous set of
instantaneous phases} $\phi_s(t)$ the continuous wavelet transform
is the convenient mathematical tool.

Let us consider continuous wavelet transform of chaotic time
series $x(t)$
\begin{equation}
W(s,t_0)=\int\limits_{-\infty}^{+\infty}x(t)\psi^*_{s,t_0}(t)\,dt,
\label{eq:WvtTrans}
\end{equation}
where $\psi_{s,t_0}(t)$ is the wavelet--function related to the
mother--wavelet $\psi_{0}(t)$ as
\begin{equation}
\psi_{s,t_0}(t)=\frac{1}{\sqrt{s}}\psi\left(\frac{t-t_0}{s}\right).
\label{eq:Wvt}
\end{equation}
The time scale $s$ corresponds to the width of the wavelet
function $\psi_{s,t_0}(t)$, and $t_0$ is shift of wavelet along
the time axis, the symbol ``$*$'' in~(\ref{eq:WvtTrans}) denotes
complex conjugation. It should be noted that the time scale $s$ is
usually used instead of the frequency $f$ of Fourier
transformation and can be considered as the quantity inversed to
it.

The Morlet--wavelet~\cite{Grossman:1984_Morlet}
\begin{equation}
\psi_0(\eta)=\frac{1}{\sqrt[4]{\pi}}\exp(j\Omega_0\eta)\exp\left(\frac{-\eta^2}{2}\right)
\label{eq:Morlet}
\end{equation}
has been used as a mother--wavelet function. The choice of
parameter value $\Omega_0=2\pi$ provides the relation ${s=1/f}$
between the time scale $s$ of wavelet transform and frequency $f$
of Fourier transformation.

The wavelet surface
\begin{equation}
W(s,t_0)=|W(s,t_0)|e^{j\phi_s(t_0)} \label{eq:WVT_Phase}
\end{equation}
describes the system's dynamics on every time scale $s$ at the
moment of time $t_0$. The value of $|W(s,t_0)|$ indicates the
presence and intensity of the time scale $s$ mode in the time
series $x(t)$ at the moment of time $t_0$. It is possible to
consider the quantities
\begin{equation}
E(s,t_0)=|W(s,t_0)|^2 \label{eq:Energy}
\end{equation}
and
\begin{equation}
\langle E(s)\rangle=\int|W(s,t_0)|^2\,dt_0, \label{eq:IntEnergy}
\end{equation}
which are instantaneous and integral energy distributions on time
scales, respectively.

At the same time, the phase $\phi_s(t)=\arg\,W(s,t)$ is naturally
introduced for every time scale $s$. It means that it is possible
to describe the behavior of each time scale $s$ by means of its
own phase $\phi_s(t)$. If two interacting chaotic oscillators are
synchronized it means that in time series $\mathbf{x}_1(t)$ and
$\mathbf{x}_2(t)$ there are scales $s$ correlated with each other.
To detect such correlation one can examine the
condition~(\ref{eq:SPhaseLocking}) which should be satisfied for
synchronized time scales.

\section{Phase synchronization of two R\"ossler systems}
\label{Sct:PSSynchro}

Let us start our consideration with two mutually coupled R\"ossler
systems with slightly mismatched
parameters~\cite{Rosenblum:1996_PhaseSynchro,Osipov:1997_PhaseSynchro}
\begin{equation}
\begin{array}{l}
\dot
x_{1,2}=-\omega_{1,2}y_{1,2}-z_{1,2}+\varepsilon(x_{2,1}-x_{1,2}),\\
\dot y_{1,2}=\omega_{1,2}x_{1,2}+ay_{1,2},\\
\dot z_{1,2}=p+z_{1,2}(x_{1,2}-c), \label{eq:Rossler}
\end{array}
\end{equation}
where $a=0.165$, $p=0.2$, and $c=10$. The parameters
$\omega_{1,2}=\omega_0\pm\Delta$ determine the parameter
mistuning, $\varepsilon$ is the coupling parameter
($\omega_0=0.97$, $\Delta=0.02$).
In~\cite{Rosenblum:1997_LagSynchro} it has been shown that for
these control parameter values and coupling parameter
$\varepsilon=0.05$ the phase synchronization is observed.

\begin{figure}
\vspace*{3cm}
\centerline{\includegraphics*[scale=0.5]{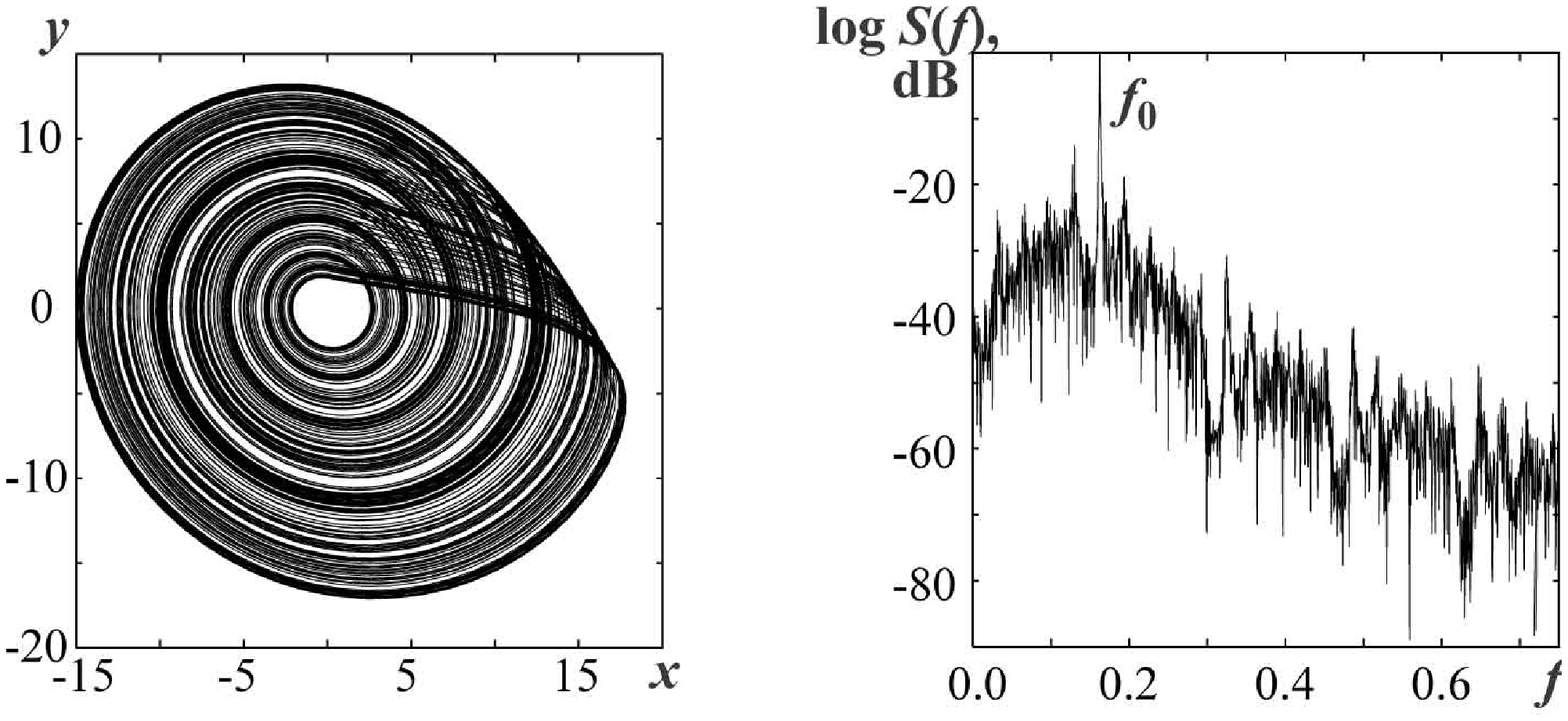}}
\centerline{\large\quad\textit{a}\qquad\qquad\qquad\qquad\qquad\quad\textit{b}}
\vspace*{1cm}\caption{\label{fgr:Rossler}} \vspace*{3cm}
\end{figure}

For this case the phase of chaotic signal can be easily introduced
in one of the ways mentioned above, because the phase coherent
attractor with rather simple topological properties is realized in
the system phase space. The attractor projection on the
$(x,y)$--plane resembles the smeared limit cycle where the phase
point always rotates around the origin
(Fig.~\ref{fgr:Rossler},\textit{a}). The Fourier spectrum $S(f)$
contains the main frequency peak $f_0\simeq 0.163$ (see
Fig.~\ref{fgr:Rossler},\textit{b}) which coincides with the mean
frequency $\bar{f}=\bar{\Omega}/2\pi$, determined from the
instantaneous phase $\phi(t)$ dynamics~(\ref{eq:MeanFrequency}).
Therefore, there are no complications to detect the phase
synchronization regime in two coupled R\"ossler
systems~(\ref{eq:Rossler}) by means of traditional approaches.

When the coupling parameter $\varepsilon$ is equal to $0.05$ the
phase synchronization between chaotic oscillators is observed. The
phase locking condition~(\ref{eq:PhaseLocking}) is satisfied and
the mean frequencies $\bar{\Omega}_{1,2}$ are entrained. So, the
time scales $s_0\simeq 6$ of both chaotic systems corresponding to
the mean frequencies $\bar{\Omega}_{1,2}$ should be correlated
with each other. Correspondingly, the phases $\phi_{s1,2}(t)$
associated with these time scales $s$ should be locked and the
condition~(\ref{eq:SPhaseLocking}) should be satisfied. The time
scales which are the nearest to the time scale $s_0$ should also
be correlated, but the interval of the correlated time scales
depends upon the coupling strength. At the same time, there should
be time scales which remain uncorrelated. These uncorrelated time
scales cause the difference between chaotic oscillations of
coupled systems.

\begin{figure}
\vspace*{3cm}
\centerline{\includegraphics*[scale=0.5]{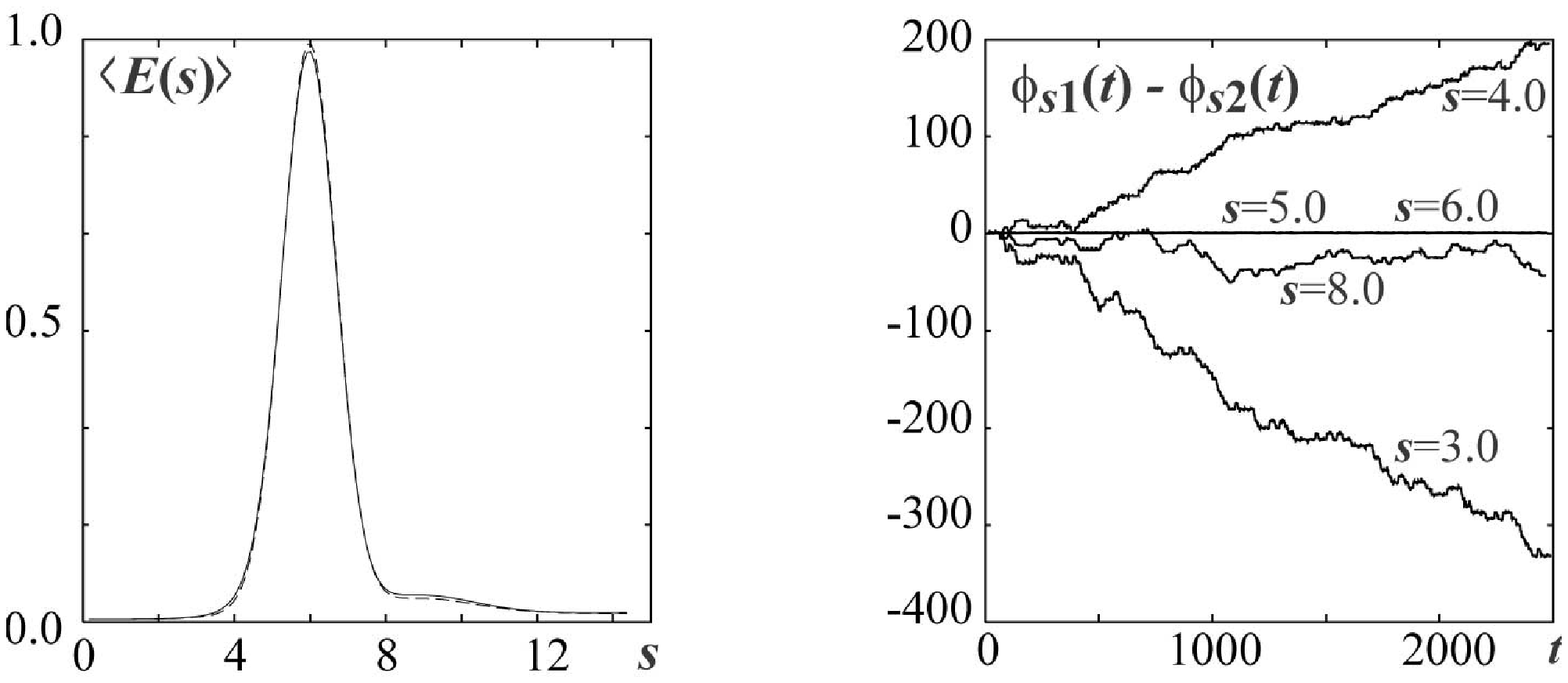}}
\centerline{\large\quad\textit{a}\qquad\qquad\qquad\qquad\qquad\quad\textit{b}}
\vspace*{1cm} \caption{\label{fgr:WVTForCoherent}} \vspace*{3cm}

\end{figure}

Figure~\ref{fgr:WVTForCoherent} illustrates the behavior of
different time scales for two coupled R\"ossler
systems~(\ref{eq:Rossler}) with phase coherent attractors. It is
clear, that the phase difference ${\phi_{s1}(t)-\phi_{s2}(t)}$ for
scales $s_0=6$ is bounded and therefore time scales $s_0=6$
corresponding to the main frequency of Fourier spectrum $f_0$ are
synchronized. It is important to note that wavelet power spectra
$\langle E_{1,2}(s)\rangle$ close to each other (see
Fig.~\ref{fgr:WVTForCoherent},\textit{a}) and time scales $s$
characterized by the large value of energy (e.g., $s$=5) close to
the main time scale $s_0=6.0$ are correlated, too. There are also
time scales which are not synchronized, like $s=3.0$, $s=4.0$,
etc. (see Fig.~\ref{fgr:WVTForCoherent},\textit{b}).

So, the phase synchronization of two mutually coupled chaotic
oscillators with phase coherent attractors manifests itself as
synchronous behavior of the time scales $s_0$ (and time scales $s$
close to $s_0$) corresponding to the chaotic signal mean frequency
$\bar{\Omega}$.

\section{Synchronization of two R\"ossler systems with funnel attractors}
\label{Sct:IllPhase}

Let us consider the more complicated example when it is impossible
to correctly introduce the instantaneous phase $\phi(t)$ of
chaotic signal $\mathbf{x}(t)$. It is clear, that for such cases
the traditional methods of the phase synchronization detecting
fail and it is necessary to use the other techniques, e.g.,
indirect measurements~\cite{Rosenblum:2002_FrequencyMeasurement}.
On the contrary, our approach gives correct results and allows to
detect the synchronization between chaotic oscillators as easily
as before.

\begin{figure}
\vspace*{3cm}
\centerline{\includegraphics*[scale=0.5]{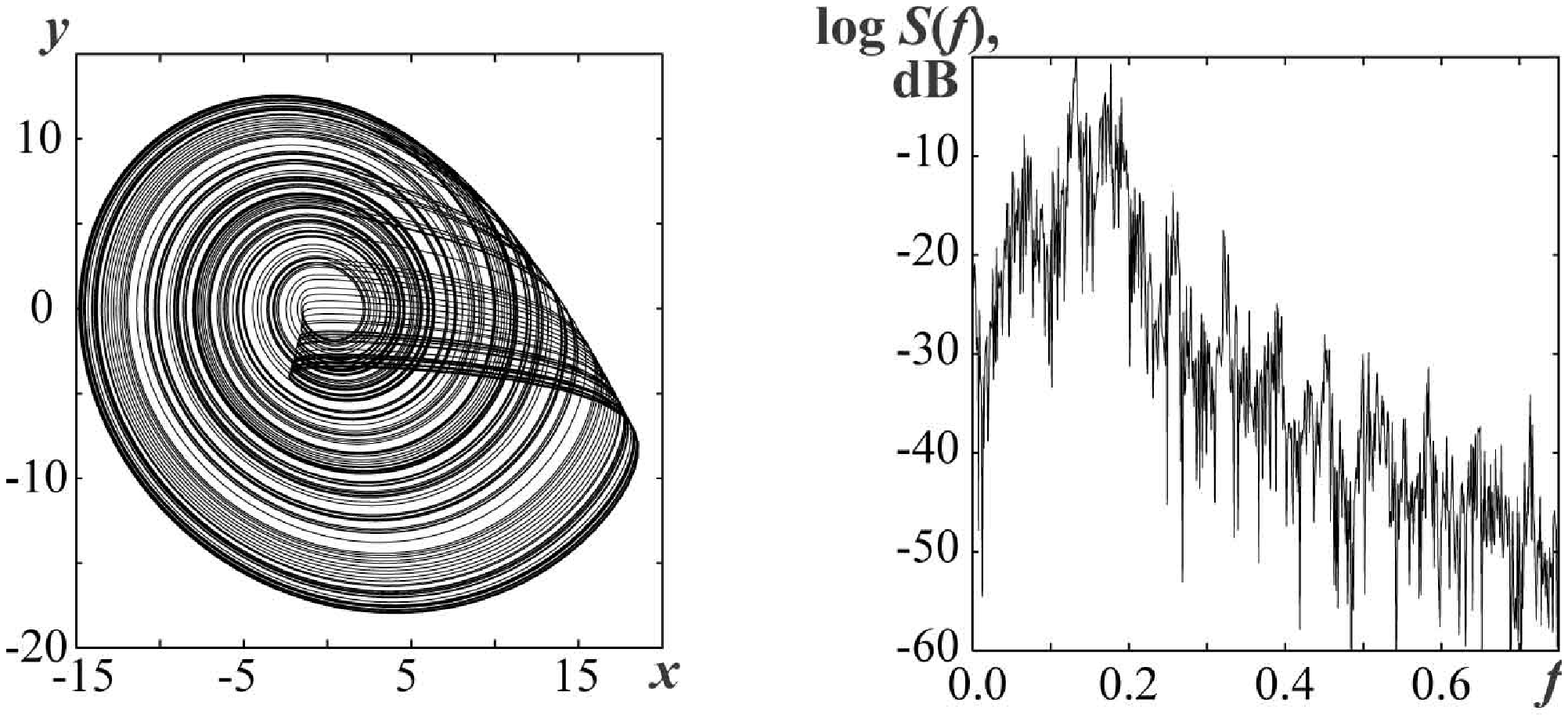}}
\centerline{\large\quad\textit{a}\qquad\qquad\qquad\qquad\qquad\quad\textit{b}}
\vspace*{1cm}\caption{ \label{fgr:FunnelRoessler}} \vspace*{3cm}
\end{figure}

To illustrate it we consider two non--identical coupled R\"ossler
systems with funnel attractors (Fig.~\ref{fgr:FunnelRoessler}):
\begin{equation}
\begin{array}{l}
\dot
x_{1,2}=-\omega_{1,2}y_{1,2}-z_{1,2}+\varepsilon(x_{2,1}-x_{1,2}),\\
\dot y_{1,2}=\omega_{1,2}x_{1,2}+ay_{1,2}+\varepsilon(y_{2,1}-y_{1,2}),\\
\dot z_{1,2}=p+z_{1,2}(x_{1,2}-c), \label{eq:FunnelRoessler}
\end{array}
\end{equation}
where $\varepsilon$ is a coupling parameter, $\omega_1=0.98$,
$\omega_2=1.03$. The control parameter values have been selected
by analogy with~\cite{Rosenblum:2002_FrequencyMeasurement} as
$a=0.22$, $p=0.1$, $c=8.5$. It is necessary to note that under
these control parameter values none of the methods mentioned above
permits to define phase of chaotic signal correctly in whole range
of coupling parameter $\varepsilon$ variation. Therefore, nobody
can determine by means of direct measurements whether the
synchronization regime takes place for several values of parameter
$\varepsilon$. On the contrary, our approach allows to detect
synchronization between considered coupled oscillators easily for
all values of coupling parameter.

\begin{figure}
\vspace*{3cm}

\centerline{\includegraphics*[scale=0.5]{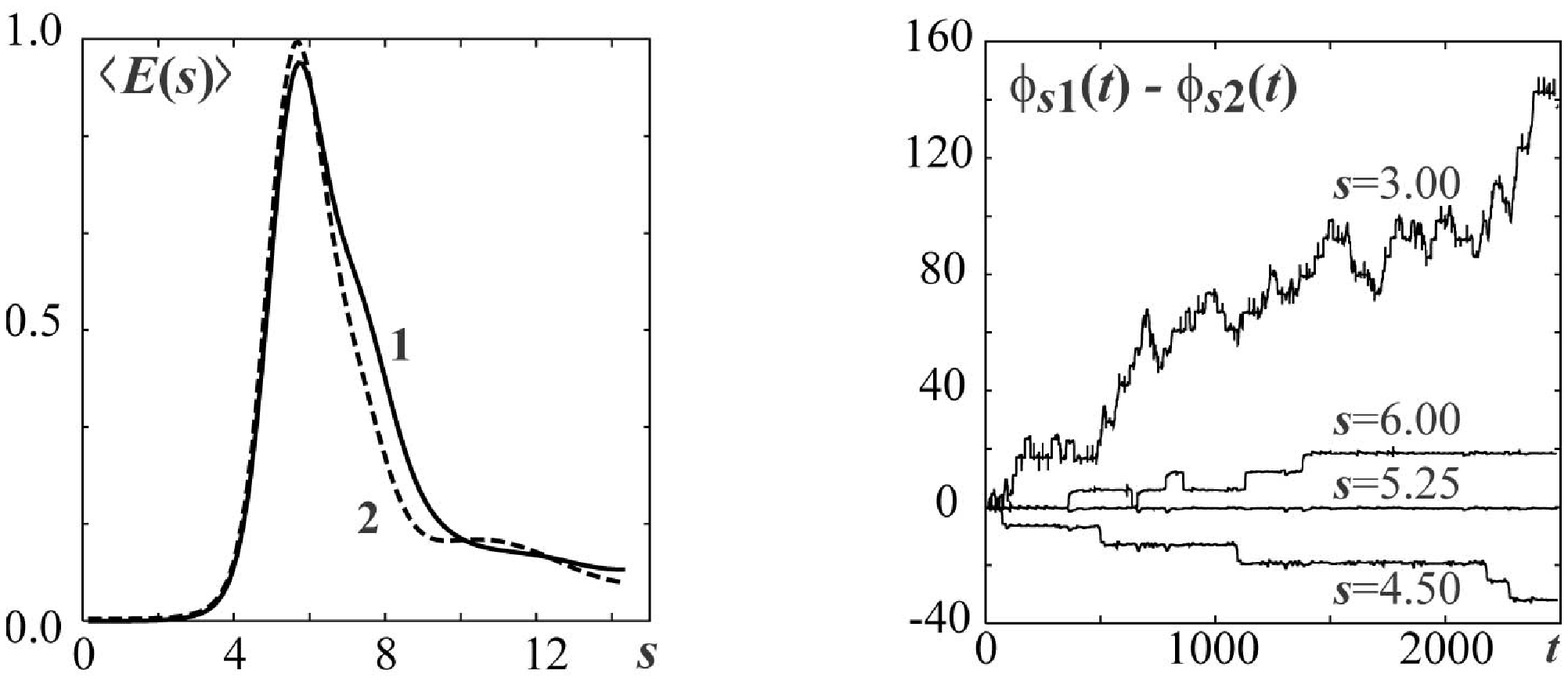}}
\centerline{\large\quad\textit{a}\qquad\qquad\qquad\qquad\qquad\quad\textit{b}}
\vspace*{1cm}\caption{\label{fgr:FunRossK=0_05}} \vspace*{3cm}

\end{figure}

In~\cite{Rosenblum:2002_FrequencyMeasurement} it has been shown by
means of the indirect measurements that for the coupling parameter
value $\varepsilon=0.05$ the synchronization of two mutually
coupled R\"ossler systems~(\ref{eq:FunnelRoessler}) takes place.
Our approach based on the analysis of the dynamics of different
time scales $s$ gives analogous results. So, the behavior of the
phase difference $\phi_{s1}(t)-\phi_{s2}(t)$ for this case has
been presented in figure~\ref{fgr:FunRossK=0_05},\textit{b}. One
can see that the phase locking takes place for the time scales
$s=5.25$ which are characterized by the largest energy value in
the wavelet power spectra $\langle E(s)\rangle$ (see
Fig.~\ref{fgr:FunRossK=0_05},\textit{a}).

It is important to note that the phase difference
$\phi_{s1}(t)-\phi_{s2}(t)$ is also bounded on the time scales
close to $s=5.25$. So, we can say that the time scales $s=5.25$
(and close to them) of two oscillators are synchronized with each
other. At the same time the other time scales (e.g., $s=4.5, 6.0$
et. al.) remain uncorrelated. For such time scales the phase
locking has not been observed (see
Fig.~\ref{fgr:FunRossK=0_05},\textit{b}).

It is clear, that the mechanism of the synchronization of coupled
chaotic oscillators is the same in both cases considered in
sections~\ref{Sct:PSSynchro} and \ref{Sct:IllPhase}. The
synchronization phenomenon is caused by the existence of time
scales $s$ in system dynamics correlated with each other.
Therefore, there is no reason to divide the considered
synchronization examples into different types.

It has been shown in~\cite{Rosenblum:1997_LagSynchro} that there
is certain relationship between PS, LS and CS for chaotic
oscillators with slightly mismatched parameters. With the increase
of coupling strength the systems undergo the transition from
unsynchronized chaotic oscillations to the phase synchronization.
With a further increase of coupling the lag synchronization is
observed. Further increasing of the coupling parameter leads to
the decreasing of the time lag and both systems tend to have the
complete synchronization regime.

\begin{figure}[tb]
\vspace*{3cm}
\centerline{\includegraphics*[scale=0.5]{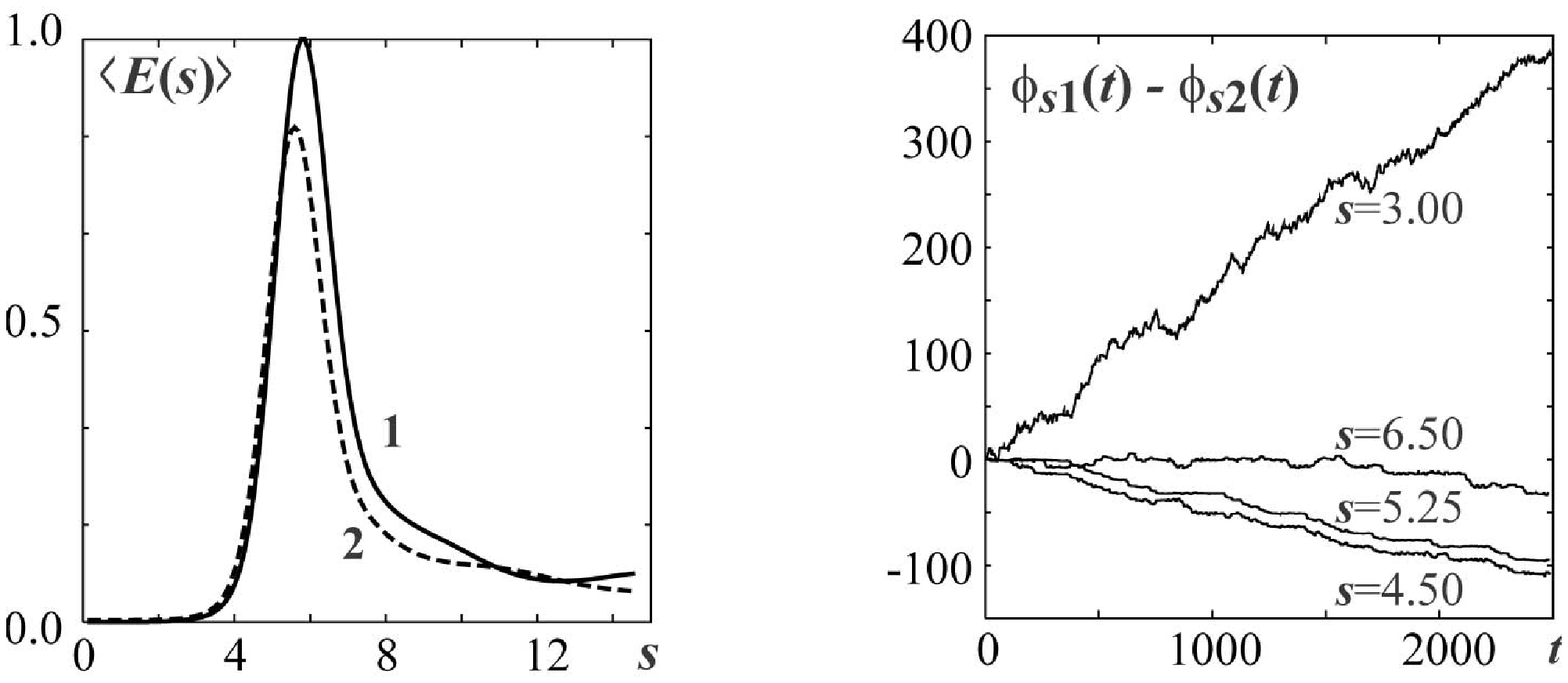}}
\centerline{\large\quad\textit{a}\qquad\qquad\qquad\qquad\qquad\quad\textit{b}}
\vspace*{1cm}\caption{\label{fgr:FunRossK=0_025}} \vspace*{3cm}
\end{figure}

Let us consider the dynamics of different time scales $s$ of two
nonidentical mutually coupled R\"ossler
systems~(\ref{eq:FunnelRoessler}) when the coupling parameter
value increases. If there is no phase synchronization between the
oscillators, then their dynamics remain uncorrelated for all time
scales $s$. Figure~\ref{fgr:FunRossK=0_025} illustrates the
dynamics of two coupled R\"ossler systems when the coupling
parameter $\varepsilon$ is small enough ($\varepsilon=0.025$). The
power spectra $\langle E(s)\rangle$ of wavelet transform for
R\"ossler systems differ from each other
(Fig.~\ref{fgr:FunRossK=0_025},\textit{a}), but the maximum values
of the energy correspond approximately to the same time scale $s$
in both systems. It is clear, that the phase difference
$\phi_{s1}(t)-\phi_{s2}(t)$ is not bounded for almost all time
scales (see Fig.~\ref{fgr:FunRossK=0_025},\textit{b}). One can see
that the phase difference $\varphi_{s1}(t)-\varphi_{s2}(t)$
increases for time scale $s=3.0$, but decreases for $s=4.5$. It
means that there should be the time scale $3<s^*<4.5$ the phase
difference on which remains bounded. This time scale $s^*$ plays a
role of a point separating the time scale areas with the phase
difference increasing and decreasing, respectively. In this case
the measure of time scales on which the phase difference remains
bounded is zero and we can not say about the synchronous behavior
of coupled chaotic oscillators (see also
section~\ref{Sct:Measure}).

As soon as any of the time scales of the first chaotic oscillator
becomes correlated with the other one of the second oscillator
(e.g., when the coupling parameter increases), the phase
synchronization occurs (see Fig.~\ref{fgr:FunRossK=0_05}). The
time scales $s$ characterized by the largest value of energy in
wavelet spectrum $\langle E(s)\rangle$ are more likely to become
correlated first. The other time scales remain uncorrelated as
before. The phase synchronization between chaotic oscillators
leads to the phase locking~(\ref{eq:SPhaseLocking}) on the
correlated time scales $s$.

\begin{figure}[tb]
\vspace*{3cm}
\centerline{\includegraphics*[scale=0.5]{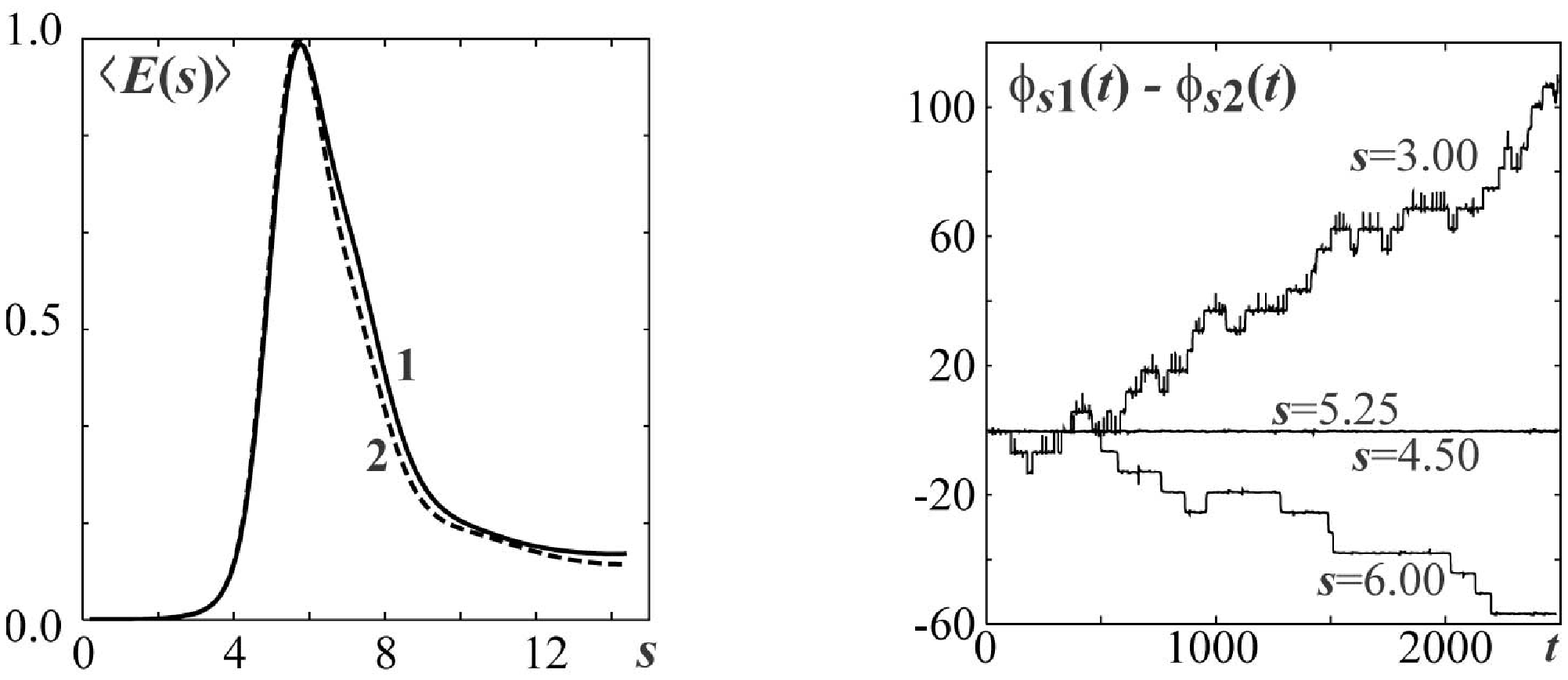}}
\centerline{\large\quad\textit{a}\qquad\qquad\qquad\qquad\qquad\quad\textit{b}}
\vspace*{1cm}\caption{\label{fgr:FunRossK=0_07}} \vspace*{3cm}
\end{figure}

When the parameter of coupling between chaotic oscillators
increases, more and more time scales become correlated and one can
say that the degree of the synchronization grows. So, with the
further increasing of the coupling parameter value (e.g.,
$\varepsilon=0.07$) in the coupled R\"ossler
systems~(\ref{eq:FunnelRoessler}) the time scales which were
uncorrelated before become synchronized (see
Fig.~\ref{fgr:FunRossK=0_07},\textit{b}). It is evident, that the
time scales $s=4.5$ are synchronized in comparison with the
previous case ($\varepsilon=0.05$,
Fig.~\ref{fgr:FunRossK=0_05},\textit{b}) when these time scales
were uncorrelated. The number of time scales $s$ demonstrating the
phase locking increases, but there are nonsynchronized time scales
as before (e.g., the time scales $s=3.0$ and $s=6.0$ remain still
nonsynchronized).

\begin{figure}[b]
\vspace*{3cm}
\centerline{\includegraphics*[scale=0.5]{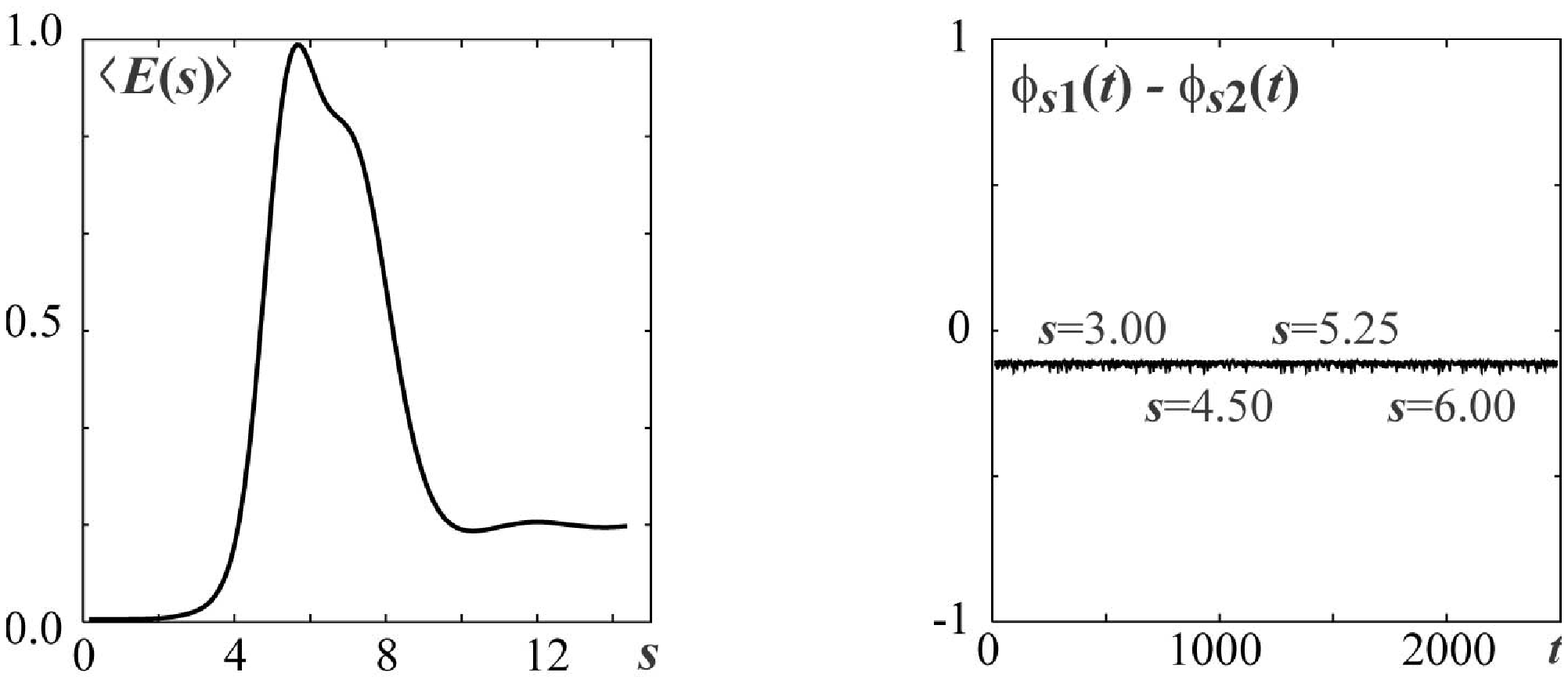}}
\centerline{\large\quad\textit{a}\qquad\qquad\qquad\qquad\qquad\quad\textit{b}}
\vspace*{1cm}\caption{\label{fgr:FunRossK=0_25}} \vspace*{3cm}
\end{figure}

Arising of the lag
synchronization~\cite{Rosenblum:1997_LagSynchro} between
oscillators means that all time scales are correlated. Indeed,
from the condition of the lag--synchronization ${x_1(t-\tau)\simeq
x_2(t)}$ one can obtain that ${W_1(s,t-\tau)\simeq W_2(t,s)}$ and
therefore ${\phi_{s1}(t-\tau)\simeq\phi_{s2}(t)}$. It is clear, in
this case the phase locking condition~(\ref{eq:SPhaseLocking}) is
satisfied for all time scales $s$. For instance, when the coupling
parameter of chaotic oscillators~(\ref{eq:FunnelRoessler}) becomes
large enough ($s=0.25$) the lag synchronization of two coupled
oscillators appears. In this case the power spectra of wavelet
transform coincide with each other (see
Fig.~\ref{fgr:FunRossK=0_25},\textit{a}) and the phase locking
takes place for all time scales $s$
(Fig.~\ref{fgr:FunRossK=0_25},\textit{b}). It is important to note
that the phase difference $\phi_{s1}(t)-\phi_{s2}(t)$ is not equal
to zero for the case of the lag synchronization. It is clear that
this difference depends on the time lag $\tau$.

Further increasing of the coupling parameter leads to the
decreasing of the time lag
$\tau$~\cite{Rosenblum:1997_LagSynchro}. Both systems tend to have
the complete synchronization regime $x_1(t)\simeq x_2(t)$, so the
phase difference $\phi_{s1}(t)-\phi_{s2}(t)$ tends to be a zero
for all time scales.

\begin{figure}
\vspace*{3cm} \centerline{\includegraphics*[scale=0.5]{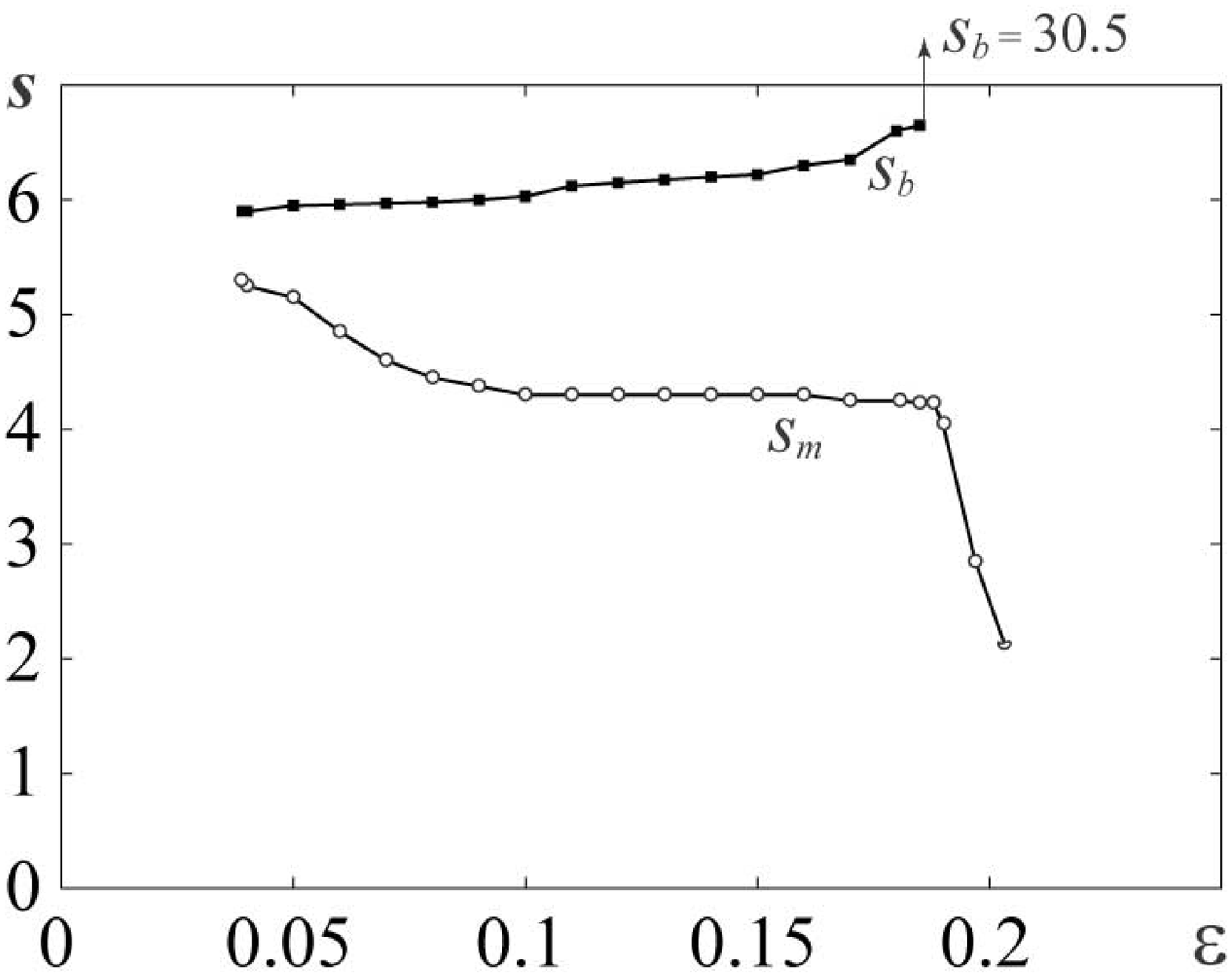}}
\vspace*{1cm}\caption{ \label{fgr:s}}\vspace*{3cm}

\end{figure}

The dependence of synchronized time scale range $[s_m;s_b]$ on
coupling parameter has been shown in Fig.~\ref{fgr:s}. The range
$[s_m;s_b]$ of synchronized time scales appears at
$\varepsilon\approx 0.039$. The appearance of synchronized time
scale range corresponds to the phase synchronization regime. When
the coupling parameter value increases the range of synchronized
time scales expands until all time scales become synchronized.
Synchronization of all time scales means the presence of lag
synchronization regime.

So, we can say the time scale synchronization (TSS) is the most
general synchronization type uniting (at least) PS, LS and CS
regimes.

\section{Generalized synchronization regime}
\label{Sct:GSSynchro}

Let us consider another type of synchronized behavior, so--called
the generalized synchronization. It has been shown above, that PS,
LS and CS are naturally interrelated with each other and the
synchronization type depends on the number of synchronized time
scales. The details of the relations between PS and GS is not at
all clear. There are several
works~\cite{Parlitz:1996_PhaseSynchroExperimental,%
Zhigang:2000_GSversusPS} dealing with the problem, how GS and PS
are correlated with each other. For instance,
in~\cite{Zhigang:2000_GSversusPS} it has been reported that two
unidirectional coupled R\"ossler systems can demonstrate the
generalized synchronization while the phase synchronization has
not been observed. This case allows to be explained easily by
means of the time scale analysis. The equations of R\"ossler
system are
\begin{equation}
\begin{array}{l}
\dot x_{1}=-\omega_{1}y_{1}-z_{1},\\
\dot y_{1}=\omega_{1}x_{1}+ay_{1},\\
\dot z_{1}=p+z_{1}(x_{1}-c)\\
\dot x_{2}=-\omega_{2}y_{2}-z_{2}+\varepsilon(x_1-x_2),\\
\dot y_{2}=\omega_{2}x_{2}+ay_{2},\\
\dot z_{2}=p+z_{2}(x_{2}-c), \label{eq:UniDirecRosslers}
\end{array}
\end{equation}
where $\mathbf{x}_1=(x_1,y_1,z_1)^T$ and
$\mathbf{x}_2=(x_2,y_2,z_2)^T$ are the state vectors of the first
(drive) and the second (response) R\"ossler systems, respectively.
The control parameter values have been chosen as $\omega_1=0.8$,
$\omega_2=1.0$, $a=0.15$, $p=0.2$, $c=10$ and $\varepsilon=0.2$.
The generalized synchronization takes place in this case
(see~\cite{Zhigang:2000_GSversusPS} for detail). Why it is
impossible to detect the phase synchronization in the
system~(\ref{eq:UniDirecRosslers}) despite the generalized
synchronization is observed becomes clear from the time scale
analysis.

\begin{figure}[tbh]
\vspace*{3cm}
\centerline{\includegraphics*[scale=0.5]{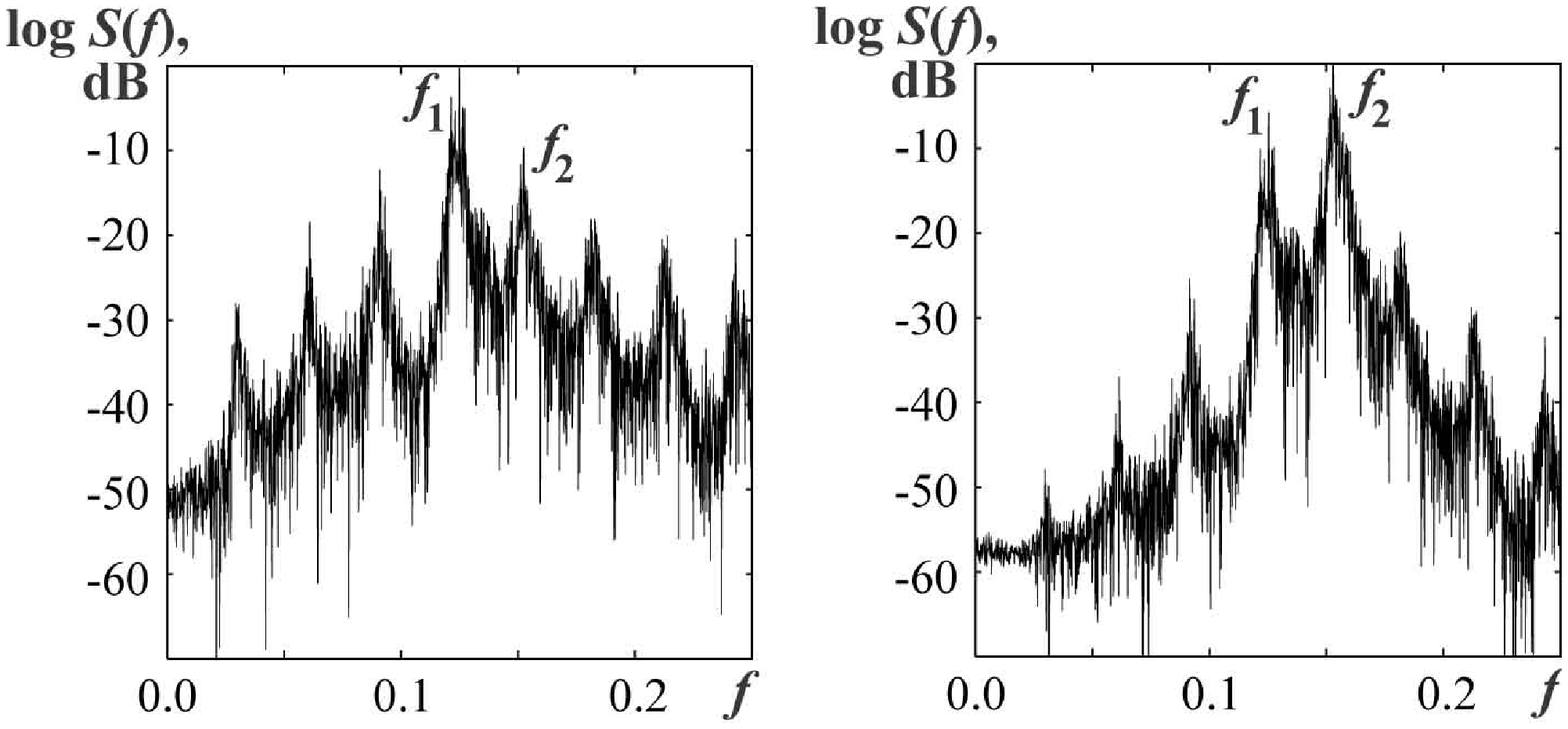}}
\centerline{\large\quad\textit{a}\qquad\qquad\qquad\qquad\qquad\quad\textit{b}}
\vspace*{1cm}\caption{ \label{fgr:GS2_spectra}} \vspace*{3cm}
\end{figure}

\begin{figure}[tbh]
\vspace*{3cm}
\centerline{\includegraphics*[scale=0.5]{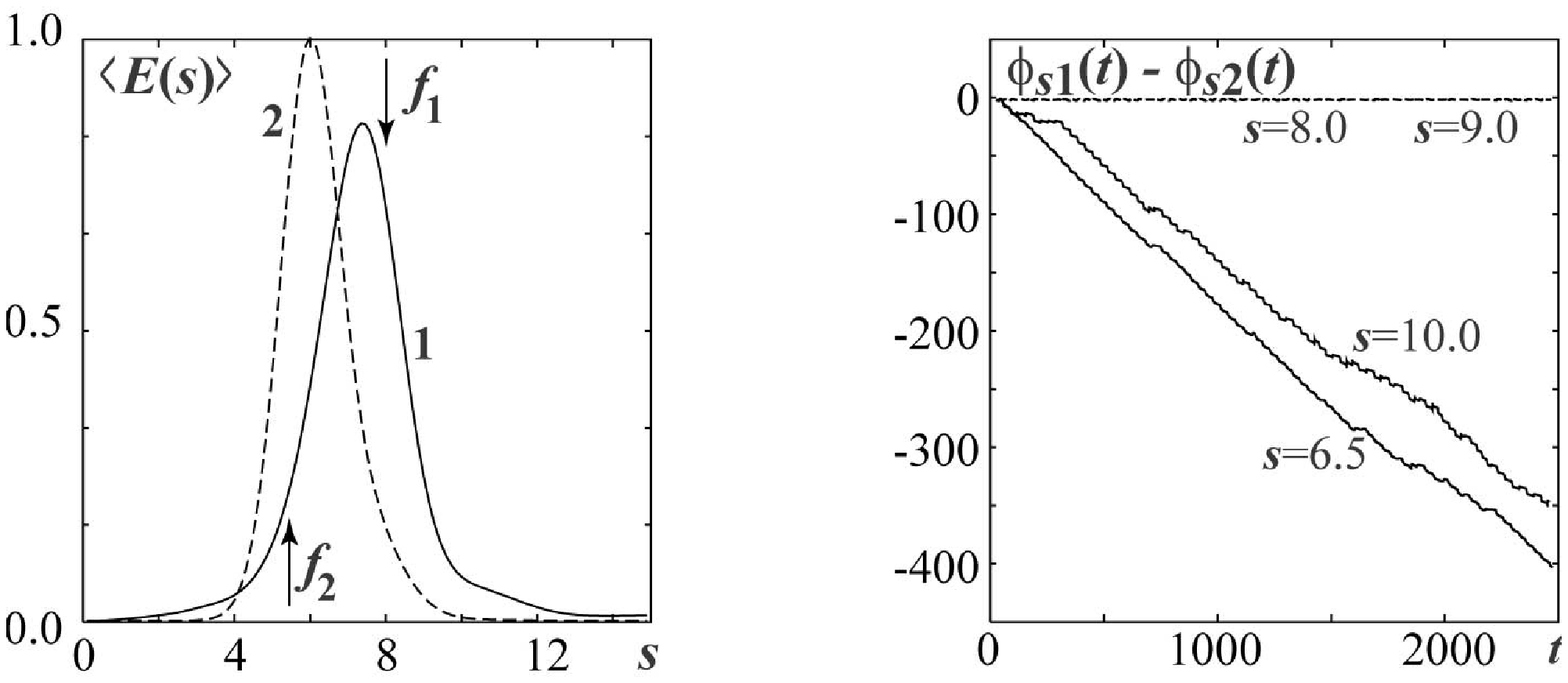}}
\centerline{\large\quad\textit{a}\qquad\qquad\qquad\qquad\qquad\quad\textit{b}}
\vspace*{1cm}\caption{ \label{fgr:GS2_wvt}} \vspace*{3cm}
\end{figure}

Let us consider Fourier spectra of coupled chaotic oscillators
(see Fig.~\ref{fgr:GS2_spectra}). There are two main spectral
components with frequencies $f_1=0.125$ and $f_2=0.154$ in these
spectra. The analysis of behavior of time scales has shown that
both the time scales $s_1=1/f_1=8$ of coupled oscillators
corresponding to the frequency $f_1$ and time scales close to
$s_1$ are synchronized while the time scales $s_2=1/f_2\simeq 6.5$
and close to them do not demonstrate synchronous behavior
(Fig.~\ref{fgr:GS2_wvt},\textit{b}).

The source of such behavior of time scales becomes clear from the
wavelet power spectra $\langle E(s)\rangle$ of both systems (see
Fig.~\ref{fgr:GS2_wvt},\textit{a}). The time scale $s_1$ of the
drive R\"ossler system is characterized by the large value of
energy while the part of energy being fallen on this scale of the
response system is quite small. Therefore, the drive system
dictates its own dynamics on the time scale $s_1$ to the response
system. The opposite situation takes place for the time scales
$s_2$ (see Fig.~\ref{fgr:GS2_wvt},\textit{a}). The drive system
can not dictate its dynamics to the response system because the
part of energy being fallen on this time scale is small in the
first R\"ossler system and large enough in the second one. So,
time scales $s_2$ are not synchronized.

Thus, the generalized synchronization of the unidirectional
coupled R\"ossler systems appears as the time scale synchronized
dynamics, as another synchronization types does before. It is also
clear, why the phase synchronization has not been observed in this
case. Fig.~\ref{fgr:GS2_spectra} shows that the instantaneous
phases $\phi_{1,2}(t)$ of chaotic signals $\mathbf{x}_{1,2}(t)$
introduced by means of traditional approaches are determined by
both frequencies $f_1$ and $f_2$, but only the spectral components
with the frequency $f_1$ are synchronized. So, the observation of
instantaneous phases $\phi_{1,2}(t)$ does not allow to detect the
phase synchronization in this case although the synchronization of
time scales takes place.

Thus, one can see that there is a close relationship between
different types of the chaotic oscillator synchronization.
According to results mentioned above we can say that PS, LS, CS
and GS are particular cases of TSS. Therefore, it is possible to
consider different types of synchronized behavior from the
universal position. Unfortunately, it is not clear, how one can
distinguish the phase synchronization. (Here we mean the phase
synchronization between chaotic oscillators takes place if the
instantaneous phase $\phi(t)$ of chaotic signal may be correctly
introduced by means of traditional approaches and the phase
locking condition~(\ref{eq:PhaseLocking}) is satisfied). and the
generalized synchronization using only the results obtained from
the analysis of the time scale dynamics.

\section{Measure of synchronization}
\label{Sct:Measure}

From examples mentioned above one can see that any type of
synchronous behavior of coupled chaotic oscillators leads to
arising of the synchronized time scales. Therefore, the measure of
synchronization can be introduced. This measure $\rho$ can be
defined as the the part of wavelet spectrum energy being fallen on
the synchronized time scales
\begin{equation}
\rho_{1,2}=\frac{1}{E_{1,2}}\int\limits_{s_m}^{s_b}\langle
E_{1,2}(s)\rangle\,ds,
\end{equation}
where $[s_m;s_b]$ is the range of time scales for which the
condition~(\ref{eq:PhaseLocking}) is satisfied and $E_{1,2}$ is a
full energy of the wavelet spectrum
\begin{equation}
E_{1,2}=\int\limits_{0}^{+\infty}\langle E_{1,2}(s)\rangle\,ds.
\end{equation}
This measure $\rho$ is 0 for the nonsynchronized oscillations and
1 for the case of complete and lag synchronization regimes. If the
phase synchronization regime is observed it takes a value between
0 and 1 depending on the part of energy being fallen on the
synchronized time scales. So, the synchronization measure $\rho$
allows not only to distinguish the synchronized and
nonsynchronized oscillations, but characterize quantitatively the
degree of TSS synchronization.

\begin{figure}[tb]
\vspace*{3cm} \centerline{\includegraphics*[scale=0.5]{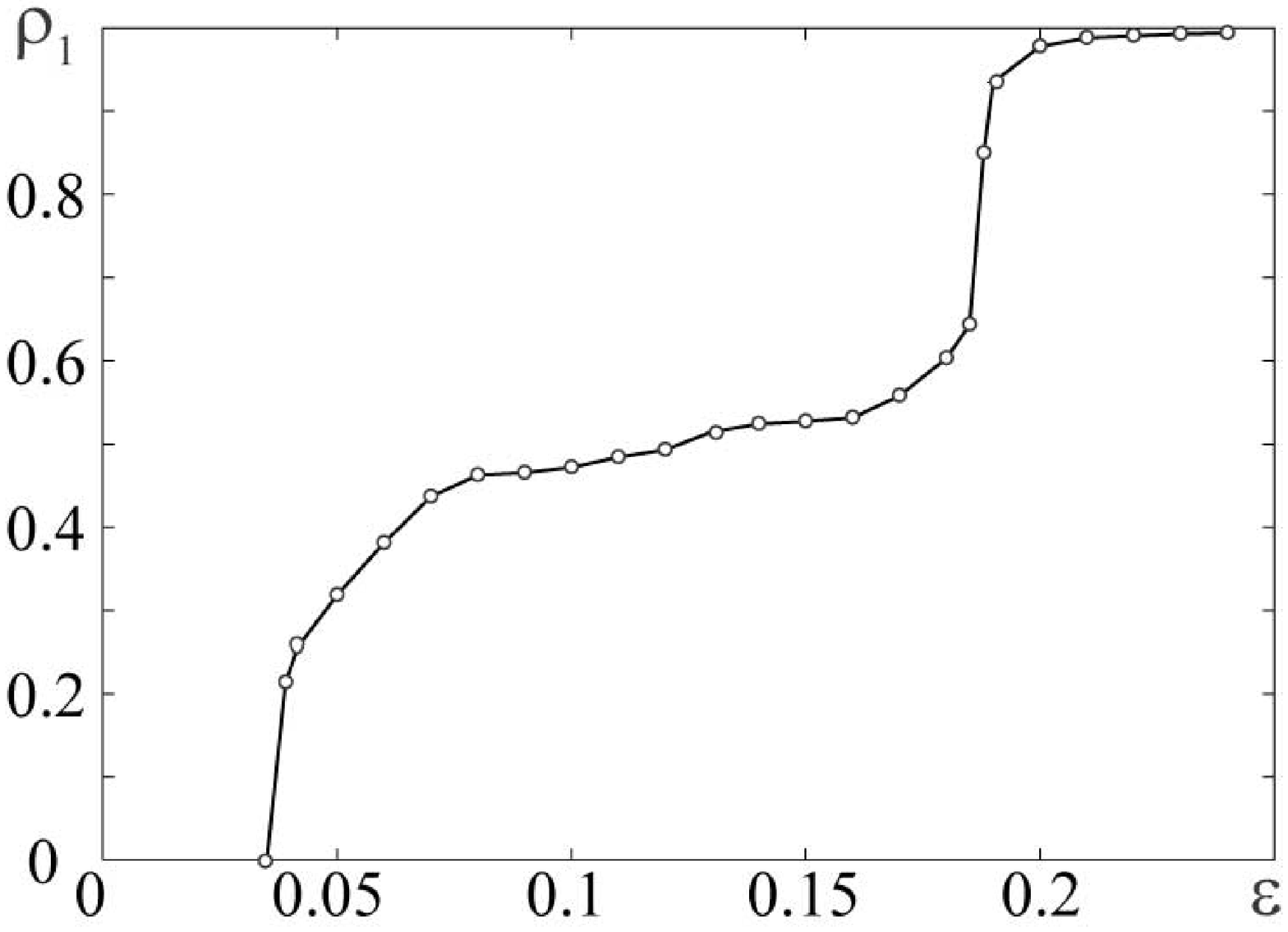}}
\vspace*{1cm}\caption{\label{fgr:rho}} \vspace*{3cm}
\end{figure}

Fig.~\ref{fgr:rho} presents the dependence of the TSS
synchronization measure $\rho_1$ for the first R\"ossler
oscillator of system~(\ref{eq:FunnelRoessler}) considered in
section~\ref{Sct:IllPhase} on the coupling parameter
$\varepsilon$. It is clear that the part of the energy being
fallen on the synchronized time scales growths monotonically with
the growth of the coupling strength. Similar results have been
obtained for the generalized synchronization of two coupled
R\"ossler systems considered in section~\ref{Sct:GSSynchro}.

It has already mentioned above that when the coupled oscillators
do not demonstrate synchronous behavior there are time scales
$s^*$ the phase difference $\varphi_{s1}(t)-\varphi_{s2}(t)$ on
which is bounded. Such time scales play role of points separating
the time scale areas where the phase difference increases and
decreases, respectively (see also section~\ref{Sct:IllPhase}).
Nevertheless, the presence of such time scales does not mean the
occurrence of chaotic synchronization because the part of energy
being fallen on them is equal to zero. Therefore, the
synchronization measure $\rho$ of such oscillations is zero, and
dynamical regime being realized in the system in this case should
be classified as non-synchronous.

\section{Conclusion}
\label{Sct:Conclusion}

Summarizing this work we would like to note several principal
aspects. Firstly, we have proposed to consider the time scale
dynamics of coupled chaotic oscillators. It allows us to consider
the different types of behavior of coupled oscillators (such as
the complete synchronization, the lag synchronization, the phase
synchronization, the generalized synchronization and the
nonsynchronized oscillations) from the universal position. In this
case TSS is the most common type of synchronous coupled chaotic
oscillator behavior. Therefore, the another types of synchronous
oscillations (PS, LS, CS and GS) may be considered as the
particular cases of TSS. The quantitative characteristic $\rho$ of
the synchronization measure has also been introduced. It is
important to note that our method (with insignificant
modifications) can also be applied to dynamical systems
synchronized by the external (e.g., harmonic) signal.

Secondly, the traditional approach for the phase synchronization
detecting based on the introducing of the instantaneous phase
$\phi(t)$ of chaotic signal is suitable and correct for such time
series characterized by the Fourier spectrum with the single main
frequency $f_0$. In this case the phase $\phi_{s0}$ associated
with the time scale $s_0$ corresponding to the main frequency
$f_0$ of the Fourier spectrum coincides approximately with the
instantaneous phase $\phi(t)$ of chaotic signal introduced by
means of the traditional approaches (see
also~\cite{Quiroga:2002_Kraskov}). Indeed, as the other
frequencies (the other time scales) do not play a significant role
in the Fourier spectrum, the phase $\phi(t)$ of chaotic signal is
close to the phase $\phi_{s0}(t)$ of the main spectral frequency
$f_0$ (and the main time scale $s_0$, respectively). It is
obvious, that in this case the mean frequencies
$\bar{f}=\langle\dot{\phi}(t)\rangle/2\pi$ and
$\bar{f}_{s0}=\langle\dot{\phi}_{s0}(t)\rangle/2\pi$ should
coincide with each other and with the main frequency $f_0$ of the
Fourier spectrum (see also~\cite{Anishchenko:2004_ChaosSynchro})
\begin{equation}
\bar{f}=\bar{f}_{s0}=f_0.
\end{equation}
If the chaotic time series is characterized by the Fourier
spectrum without the main single frequency (like the spectrum
shown in the Fig.~\ref{fgr:FunnelRoessler},\textit{b}) the
traditional approaches fail. One has to consider the dynamics of
the system on all time scales, but it is impossible to do it by
means of the instantaneous phase $\phi(t)$. On the contrary, our
approach based on the time scale dynamics analysis can be used for
both types of chaotic signals.

Finally, our approach can be easily applied to the experimental
data because it does not require any \hbox{a-priori} information
about the considered dynamical systems. Moreover, in several cases
the influence of the noise can be reduced by means of the wavelet
transform (for detail,
see~\cite{alkor:2003_WVTBookEng,Torrence:1998_Wvt,Gusev:2003_wvt}).
We believe that our approach will be useful and effective for the
analysis of physical, biological, physiological and other data,
such as~\cite{Elson:1998_NeronSynchro,Quiroga:2002_Kraskov,%
Lachaux:2000_WVTSynchro}.

\section*{Acknowledgments}
\label{Sct:Acknowledgments} We express our appreciation to
Corresponding Member of Russian Academy of Science, Professor
Dmitriy I. Trubetskov and Professors Vadim~S.~Anishchenko and
Tatyana E. Vadivasova for valuable discussions. We also thank
Svetlana V. Eremina for the support.

This work has been supported by U.S.~Civilian Research \&
Development Foundation for the Independent States of the Former
Soviet Union (CRDF), grant {REC--006} and Russian Fundation for
Basic Research, Grant 02--02--16351. A.E.H. also thanks
``Dynastiya'' Foundation.
%\end{acknowledgments}


%\bibliography{thesis}

\newpage

\centerline{\bf CAPTIONS}

{\bf Fig.\,1.} (\textit{a}) Phase coherent attractor and
(\textit{b}) spectrum of the first R\"ossler
system~(\ref{eq:Rossler}). Coupling parameter $\varepsilon$
between oscillators is equal to zero

\bigskip

{\bf Fig.\,2.} (\textit{a}) Wavelet power spectrum $\langle
E(s)\rangle$ for the first (solid line) and the second (dashed
line) R\"ossler systems~(\ref{eq:Rossler}). (\textit{b}) The
dependence of phase difference $\phi_{s1}(t)-\phi_{s2}(t)$ on time
$t$ for different time scales $s$. The coupling parameter between
oscillators is $\varepsilon=0.05$. The phase synchronization for
two coupled chaotic oscillators is observed

\bigskip

{\bf Fig.\,3.} (\textit{a}) Phase picture and (\textit{b}) power
spectrum of the first R\"ossler system~(\ref{eq:FunnelRoessler})
oscillations. Coupling parameter $\varepsilon$ is equal to zero

\bigskip

{\bf Fig.\,4.} (\textit{a}) The normalized energy distribution in
wavelet spectrum $\langle E(s)\rangle$ for the first (the solid
line denoted ``1'') and the second (the dashed line denoted ``2'')
R\"ossler systems~(\ref{eq:FunnelRoessler}); (\textit{b}) the
phase difference $\phi_{s1}(t)-\phi_{s2}(t)$ for two coupled
R\"ossler systems. The value of coupling parameter has been
selected as $\varepsilon=0.05$. The time scales $s=5.25$ are
correlated with each other and the synchronization has been
observed

\bigskip

{\bf Fig.\,5.} (\textit{a}) The normalized energy distribution in
wavelet spectrum $\langle E(s)\rangle$ for the first (the solid
line denoted ``1'') and the second (the dashed line denoted ``2'')
R\"ossler systems; (\textit{b}) the phase difference
$\phi_{s1}(t)-\phi_{s2}(t)$ for two coupled R\"ossler systems. The
value of coupling parameter has been selected as
$\varepsilon=0.025$. There is no phase synchronization between
systems

\bigskip

{\bf Fig.\,6.} (\textit{a}) The normalized energy distribution in
wavelet spectrum $\langle E(s)\rangle$ for the first (the solid
line denoted ``1'') and the second (the dashed line denoted ``2'')
R\"ossler systems; (\textit{b}) the phase difference
$\phi_{s1}(t)-\phi_{s2}(t)$ for two coupled R\"ossler systems. The
value of coupling parameter has been selected as
$\varepsilon=0.07$.

\bigskip

{\bf Fig.\,7.} (\textit{a}) The normalized energy distribution in
wavelet spectrum $\langle E(s)\rangle$ for the R\"ossler system;
(\textit{b}) the phase difference $\phi_{s1}(t)-\phi_{s2}(t)$ for
two coupled R\"ossler systems. The value of coupling parameter has
been selected as $\varepsilon=0.25$. The lag synchronization has
been observed, all time scales are synchronized

\bigskip

{\bf Fig.\,8.} The dependence of the synchronized time scale range
$[s_m;s_b]$ on the coupling strength $\varepsilon$ for two
mutually coupled R\"ossler systems~(\ref{eq:FunnelRoessler}) with
funnel attractors

\bigskip

{\bf Fig.\,9.} Fourier spectra for(\textit{a}) the first (drive)
and (\textit{b}) the second (response) R\"osler
systems~(\ref{eq:UniDirecRosslers}). The coupling parameter is
$\varepsilon=0.2$. The generalized synchronization takes place

\bigskip

{\bf Fig.\,10.} (\textit{a}) The normalized energy distribution in
wavelet spectrum $\langle E(s)\rangle$ for the first (the solid
line denoted ``1'') and the second (the dashed line denoted ``2'')
R\"ossler systems. The time scales pointed with arrows correspond
to the frequencies $f_1=0.125$ and $f_2=0.154$, respectively;
(\textit{b}) the phase difference $\phi_{s1}(t)-\phi_{s2}(t)$ for
two coupled R\"ossler systems. The generalized synchronization has
been observed

\bigskip

{\bf Fig.\,11.} The dependence of the synchronization measure
$\rho_1$ for the first R\"ossler system~(\ref{eq:FunnelRoessler})
on the coupling strength $\varepsilon$. The measure $\rho_2$ for
the second R\"ossler oscillator behaves in a similar manner, so it
has not been shown in the figure

\newpage

\thispagestyle{empty}

\newpage

\thispagestyle{empty}

\newpage

\thispagestyle{empty}

\newpage
\thispagestyle{empty}

\newpage
\thispagestyle{empty}

\newpage
\thispagestyle{empty}

\newpage
\thispagestyle{empty}

\newpage
\thispagestyle{empty}

\newpage
\thispagestyle{empty}

\newpage
\thispagestyle{empty}

\newpage
\thispagestyle{empty}

\end{document}